\documentclass[12pt]{article}
\usepackage{curves}
\usepackage{eepic}
\usepackage{epsfig}

\newcommand{\dd}{\mbox{\rm d}}

\newcommand{\ci}{\cite}

\newcommand{\be}{\begin{equation}}
\newcommand{\ee}{\end{equation}}
\newcommand{\lbl}{\label}
\newcommand{\bi}{\bibitem}

\newcommand{\vs}{\vspace}
\newcommand{\hs}{\hspace}

\def\bear{\begin{eqnarray}}
\def\ear{\end{eqnarray}}

\begin{document}
\thispagestyle{empty}


\begin{center}

{\bf \Large Towards a New Paradigm: Relativity in

Configuration Space}

\vs{6mm}

M. Pav\v si\v c

Jo\v zef Stefan Institute, Jamova 39, SI-1000 Ljubljana, Slovenia;

E-mail: matej.pavsic@ijs.si

\vs{1.5cm}

ABSTRACT
\end{center}

We consider the possibility that the basic space of physics is
not spacetime, but configuration space. We illustrate this on the
example with a system of gravitationally interacting point
particles. It turns out that such system can be described by the minimal length
action in a multidimensional configuration space ${\cal C}$ with
a block diagonal metric. Allowing for more general metrics
and curvatures of ${\cal C}$, we step beyond the ordinary general
relativity in spacetime. The latter theory is then an
approximation to the general relativity in ${\cal C}$. Other sorts
of configuration spaces can also be considered, for instance
those associated with extended objects, such as strings and branes.
This enables a deeper understanding of the geometric principle
behind string theory, and an insight on the occurrence of Yang-Mills
and gravitational fields at the `fundamental level'. 

\section{Introduction}

After many decades of intensive research there is still no general consensus on the
major persisting puzzles such as the unification of fundamental interactions,
quantum gravity, the problem of time, the cosmological constant problem,
the nature of dark matter and energy, etc. From history we know that such
situation calls for `paradigm shift'.  We also know that often a formalism
is more powerful than initially envisaged. For example, in the Hamilton-Jacobi function
there is a hint of quantum mechanics, which could have been guessed much
earlier before its experimental discovery. The line element in Minkowski
spacetime suggested its generalization to curved spacetime and thus the
theory of gravity. Clifford algebra led to the Dirac theory of electron. In all those
cases the formalism itself pointed to its own generalization! 
This introduced important new physics.

Having in mind such lessons from history it seems reasonable to do
something analogous with the currently available formalisms, and
to step beyond the existing paradigm. We will first examine the formalism
that describes a system of point particles in the presence of gravity. We will
then consider a generalization of the theory of relativity in which
spacetime $M_4$ is replaced by the configuration space ${\cal C}$
associated with a given physical system.
The system will be considered as a point that traces a geodetic
line in configuration space. Such theory predicts in general a different
dynamical behavior of a many particle system than does the ordinary theory.
But in particular, for a suitable metric of ${\cal C}$,  we obtain the ordinary
many particle action in the presence of gravitational field. In general,
the configuration space can have non vanishing curvature. From the point of
view of 4-dimensional spacetime, which is a subspace of ${\cal C}$, there
exist extra forces that act on a particle, besides the ordinary gravity.
Observations suggest that the ordinary theory of gravity cannot be straightforwardly
applied to large scale systems, such as galaxies, clusters of galaxies,
and the universe. Instead, one has to introduce the concept of dark
matter and dark energy\,\ci{dark}, or alternatively,
to consider suitable modifications of the theory of gravity
(MOND)\,\ci{mond}. We propose to explore the possibility that
general relativity, not in spacetime $M_4$, but in multidimensional
configuration space ${\cal C}$ might solve such astrophysical puzzles.
The theory can also be applied to other kinds of configuration spaces,
e.g., those associated with extended objects such as strings\ci{strings}
and branes\,\ci{branes}.
This enables a deeper understanding of the geometric principle behind the
string theory, and the insight on the occurrence of
the Yang-Mills and gravitational fields.

\section{Generalizing relativity}

\subsection{Configuration space replaces spacetime}

Let us consider a system of point particles in the presence of a gravitational field
$g_{\mu \nu}$. The action is the sum of the individual point particle
actions:
\be
I[\dot X_i ^\mu  ]\, = \sum\limits_i {} \int {d\tau \,} [\dot X_i ^\mu  \dot X_i ^\nu  
\,m_i \,g_{\mu \nu } (X_i ^\mu  )]^{1/2}
\lbl{2.1}
\ee
We will now rewrite this into an equivalent form.

Let us recall that a point particle action
\be
I[X^\mu  ]\, = \,\,\int_{}^{} {{\rm{d}}\tau \,m\,(\dot X^\mu  } \dot X^\nu  
\,g_{\mu \nu } )^{1/2} 
\lbl{2.1a}
\ee
has its equivalent in the Schild action
\be
I[X^\mu  ]\, = \,\,\int_{}^{} {{\rm{d}}\tau \,\frac{m}{k}\,\dot X^\mu  } \dot X^\nu  
\,g_{\mu \nu }
\lbl{2.2} 
\ee
which is a gauge fixed action with
\be
\dot X^\mu  \dot X^\nu  g_{\mu \nu } \, = \,k^2 = {\rm constant}
\lbl{2.3}
\ee
where $k$ is a constant.\footnote{
Variation of the action (\ref{2.1a}) gives 
$\frac{{\rm{d}}}{{{\rm{d}}\tau }}\left( {\dot X^\nu  \,g_{\mu \nu } } \right)\, 
- \,\frac{1}{2}\,g_{\alpha \beta ,\mu } \,\dot X^\alpha  \dot X^\beta  \, = \,0$.
This can be rewritten into the forms
$\frac{1}{{\sqrt {\dot X^2 } }}\frac{{\rm{d}}}{{{\rm{d}}\tau }}
\left( {\frac{{\dot X^\nu  \,g_{\mu \nu } }}{{\sqrt {\dot X^2 } }}} \right)\, 
- \,\frac{1}{2}\,g_{\alpha \beta ,\mu } \,\frac{{\dot X^\alpha  
\dot X^\beta  }}{{\dot X^2 }}\,\, - 
\frac{1}{{\sqrt {\dot X^2 } }}\frac{{\rm{d}}}{{{\rm{d}}\tau }}
\left( {\frac{1}{{\sqrt {\dot X^2 } }}} \right)\,\dot X^\nu  g_{\mu \nu }  = 
\,0$. If we multiply this by ${\dot X}^\mu$
(and sum over $\mu$), then the first two terms give identically zero, so that
we find $\sqrt{{\dot X}^2}\frac{{\rm{d}}}{{{\rm{d}}\tau }}\left( 
{\frac{1}{{\sqrt {\dot X^2 } }}} \right)\,=0$, or  ${\dot X}^2 \equiv g_{\mu \nu}
{\dot X}^\mu {\dot X}^\nu = C^2$, with $C$ being a constant. On the other hand,
the momentum belonging to the Schild action is
$p_\mu = \frac{m {\dot X}_\mu}{k}$. Thus $m^2 = p_\mu p^\mu =
\frac{m^2 {\dot X}_\mu {\dot X}^\mu}{k^2}  = \frac{m^2 C^2}{k^2}$
which implies $C^2 = k^2$. }

The Schild action for a system of point particles is
\be
I[\dot X_i ^\mu  ]\, = \int {{\rm{d}}\tau } \sum\limits_i^{} 
{\dot X_i ^\mu  \dot X_i ^\nu  \,\frac{{m_i }}{{k_i }}\,g_{\mu \nu } (X_i ^\mu  )}
\lbl{2.4} 
\ee
This can be considered as a quadratic form in a multidimensional space ${\cal C}$ whose
dimension is 4 times the number $N$ of particles in the system. To see
this more clearly, it is convenient to introduce a more compact notation:
\be
 \dot X_i ^\mu  \, \equiv \dot X^{(i\mu )} \, \equiv \,\dot X^M \,,\,\,\,\,\,\,\,M\, = \,(i\mu )
 \lbl{2.5a}
\ee
\be
 \frac{m_i}{k_i }\, g_{\mu \nu } \,\, \equiv \,\,\frac{M}{K}\, g_{({i\mu })(j\nu )} 
 \, \equiv \,\,\frac{M}{K} \, g_{MN}
\lbl{2.5b}
\ee
Then the action (\ref{2.4}) becomes
\be
I[X^M ]\, = \,\,\int {{\rm{d}}\tau } \,\,\dot X^M \dot X^N \,\frac{M}{K} \, g_{MN} (X^M )
\lbl{2.6}
\ee
which is the Schild action in ${\cal C}$. The $4N$-dimensional space ${\cal C}$
is {\it the configuration space} associated with a system.
From the context it should be clear when $M$ is a double index $M\equiv (i \mu)$, and
when it is a constant, analogous to single particle mass $m$.

From the
equations of motion derived from the action (\ref{2.6}) it follows
\be
    {\dot X}^M {\dot X}^M g_{MN} = K^2
\lbl{2.7}
\ee
where $K$ is a constant. Explicitly this reads
\be
   {\dot X}_1^2 + {\dot X}_2^2 + ... + {\dot X}_N^2 = K^2
\ee   

Rewriting the latter equation as
\be
     \frac{{\dot X}_1^2}{K^2} = 1 - \frac{{\dot X}_2^2}{K^2}  -  \frac{{\dot X}_3^2}{K^2}
     - ... - \frac{{\dot X}_N^2}{K^2}
\lbl{2.8a}
\ee
multiplying it by $M^2$, and using the expression
\be
      p_M = \frac{M {\dot X}_M}{K} \equiv \frac{M {\dot X}_{i \mu}}{K}
\lbl{2.8b}
\ee
we find
\be
      \frac{M^2 {\dot X_1}^2}{K^2} = M^2 - p_2^2 -–p_3^2 - ...…- p_N^2 = p_1^2 \equiv m_1^2
\lbl{2.9}
\ee
\be
         \frac{M^2}{K^2}=\frac{m_1^2}{k_1^2}  ~~~~~~~ {\rm or} ~~~ 
         \frac{M}{K}=\frac{m_1}{k_1}
\lbl{2.10}
\ee
where
\be
   k_i^2 = \dot X_i ^2 \, = \,g_{\left( {i\mu } \right)(i\nu )} \,\dot X_i ^\mu 
   \dot X_i ^\nu  \,\,\,\,\,\  ({\rm no~sum~over}) \, i, ~~~ i = 1,2,..., N
\lbl{2.10a}
\ee
Since the above derivation can be repeated for any $i=1,2,…,N$, we have 
\be
\frac{M}{K} = \frac{m_i}{k_i}
\lbl{2.11}
\ee
Eqs.\,(\ref{2.7}),(\ref{2.11}) and (\ref{2.11}) thus imply
\be
     M^2 = \sum_i p_i^2
\lbl{2.12}
\ee
\be
p_{i\mu }  = \frac{{M\dot X_{i\mu } }}{{\sqrt {\dot X^2 } }}\, = \,\frac{{m_i \dot X_{i\mu } }}{{\sqrt {\dot X_i ^2 } }}
\lbl{2.12a}
\ee

The Schild action in ${\cal C}$ is equivalent to the reparametrization
invariant action ${\cal C}$:
\be
I[X^M ]\, = \,M\int {{\rm{d}}\tau } \,\,[\dot X^M \dot X^N g_{MN} (X^M )]^{1/2}
\lbl{2.13}
\ee
which is proportional to the length of a worldline in ${\cal C}$. Constant $M$ has the
role of {\it mass} in ${\cal C}$.

Having arrived at the action (\ref{2.13}), we will now assume that the metric
$g_{MN}$ need not be of the block diagonal form (\ref{2.5b}).
We will assume that configuration space ${\cal C}$ is a manifold
equipped with metric $G_{MN}$, connection and curvature (that in
general does not vanish).

In particular, for the block diagonal metric
 \be
G_{MN} \, \equiv \,G_{(i\mu )(j\nu )}  = \,g_{(i\mu )(j\nu )}  = \,\left( {\begin{array}{*{20}c}
   {g_{\mu \nu } (x_1 )} & 0 & 0 & {...}  \\
   0 & {g_{\mu \nu } (x_2 )} & 0 & {...}  \\
   0 & 0 & {g_{\mu \nu } (x_3 )} & {...}  \\
    \vdots  &  \vdots  &  \vdots  &  \vdots   \\
\end{array}} \right)
\lbl{2.14}
\ee
we obtain the ordinary relativistic theory for a many particle system in
a given gravitational field.

By allowing for a more general metric, that cannot be transformed into the
form (\ref{2.14}) by a choice of coordinates in ${\cal C}$,
we go beyond the ordinary theory.

Configuration space ${\cal C}$ is the space of possible ``instantaneous" 
configurations in $M_4$. Its points are described by coordinates 
$x^M \equiv x_i^\mu$. A given configuration traces a worldline 
$x^M = X^M (\tau)$ in ${\cal C}$ (see Fig.\,1).

\setlength{\unitlength}{.8mm}

\begin{figure}[ht]
\hs{3mm} \begin{picture}(120,85)(25,0)
\put(60,2){\epsfig{file=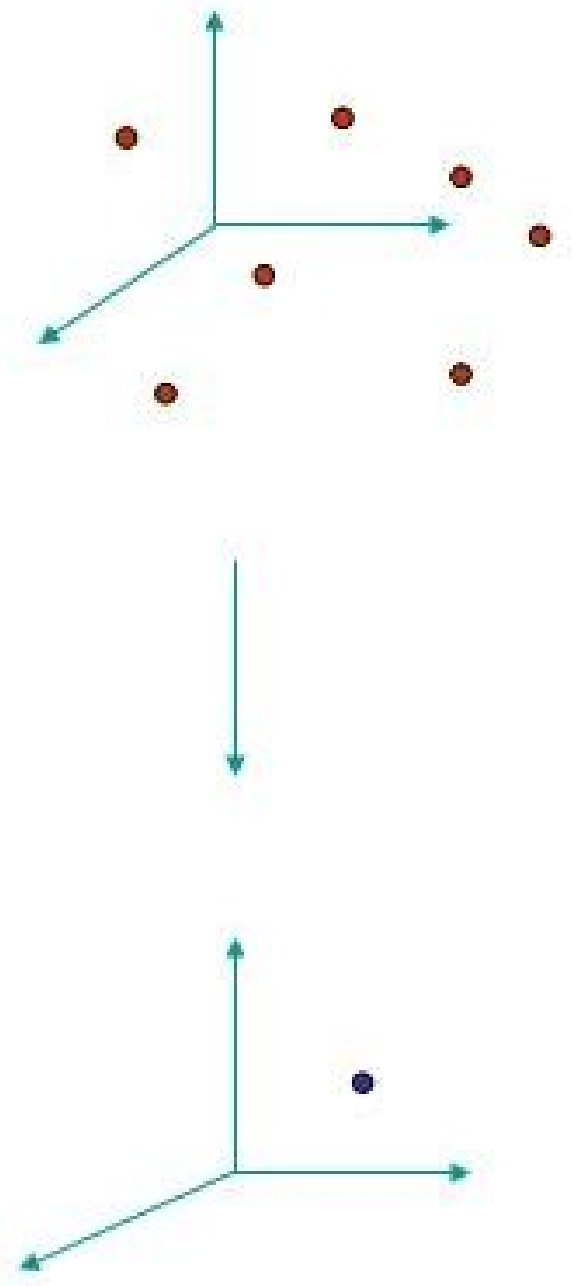,width=24.67mm}}
\put(135,-3){\epsfig{file=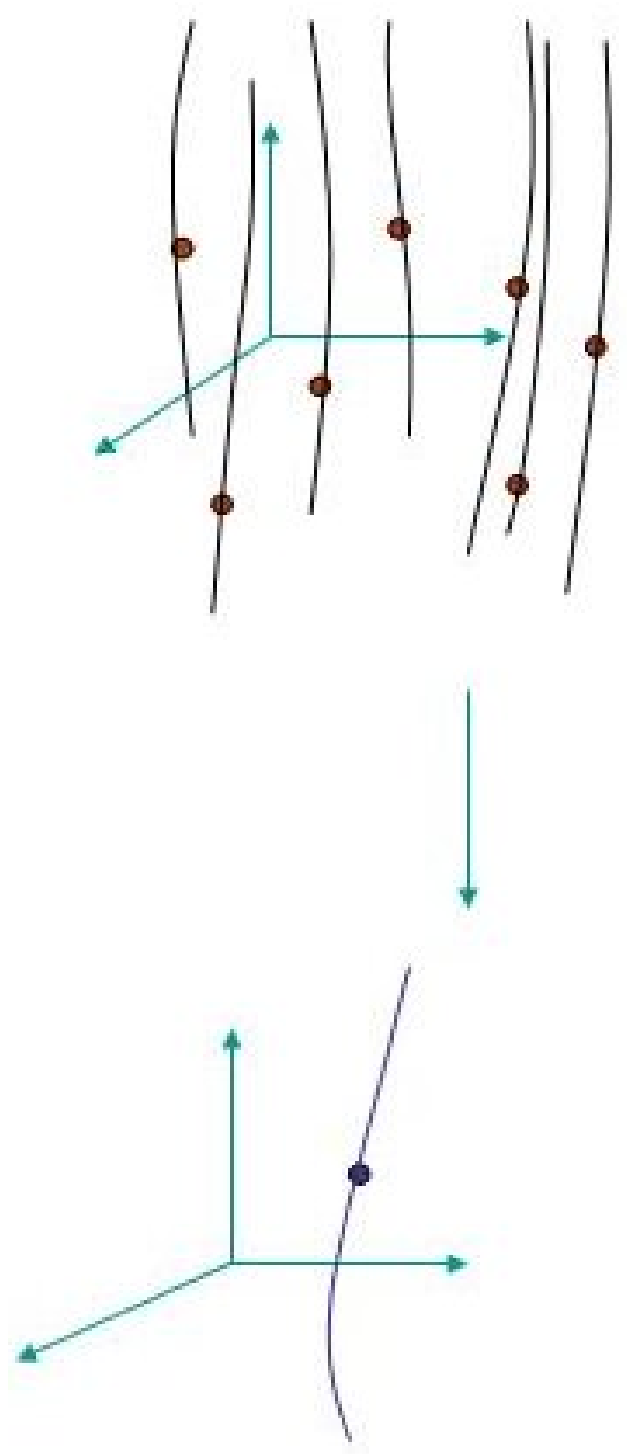,width=27.69mm}}

\put(34,80){\small `Instantaneous' configuration in $M_4$}
\put(57,69){$M_4$}
\put(63,18){${\cal C}$}
\put(87,49){$X_i^\mu$}
\put(80,15){$X^M$}

\put(120,80){\small `Evolution' of configuration in $M_4$}
\put(75,33){\small Representation in configuration space ${\cal C}$}
\put(132,69){$M_4$}
\put(169,49){$X_i^\mu(\tau)$}
\put(157,14){$X^M(\tau)$}
\put(136,18){${\cal C}$}

\end{picture}

\caption{\small An `instantaneous' configuration can be represented as a set of points in
spacetime $M_4$, or as a point in configuration space ${\cal C}$. Analogously,
a `moving' configuration can be represented as a set of worldines in $M_4$,
or a single worldline in ${\cal C}$.}
\end{figure} 

A dynamically possible worldline in ${\cal C}$ is a geodesic in ${\cal C}$,
and it satisfies the variation principle based on the action (\ref{2.13}).

Instead of considering a fixed metric of ${\cal C}$, let us assume that the
metric $G_{MN}$ is dynamical, so that the total action contains a kinetic term
for $G_{MN}$:
\be
I[X^M ,G_{MN} ] = \,I_m  + I_g 
\lbl{2.15}
\ee
where
\be
I_m \, = \int {{\rm{d}}\tau } \,\,M\,(G_{MN} \,\dot X^M \dot X^N )^{(1/2)} \, = 
\,\int {{\rm{d}}\tau } \,\,M\,(G_{MN} \,\dot X^M \dot X^N )^{(1/2)} 
\,\delta ^D \,(x\, - \,X(\tau ))\,{\rm{d}}^D x
\lbl{2.16}
\ee
and
\be
I_g \, = \,\frac{1}{{16\pi G_D }}\int {d^D x\,\sqrt {|G|} } \,{\cal R}
\lbl{2.16a}
\ee
Here ${\cal R}$ is the curvature scalar in ${\cal C}$.
So we have {\it general relativity in configuration space} ${\cal C}$.
We have arrived at a theory which is analogous to Kaluza-Klein theory.
Configuration space is a higher dimensional space, whereas
spacetime $M_4$ is a 4-dimensional subspace of ${\cal C}$,
associated with a chosen particle.

The concept of configuration space can be used either in 
macrophysics or in microphysics. Configuration space
associated with a system of point particles is finite
dimensional. Later we will discuss infinite dimensional
configuration spaces associated with strings and branes.

\subsection{Equations of motion for a configuration
of point particles}

The equations of motion derived from the action (\ref{2.15})
are the Einstein equations in configuration space ${\cal C}$.
Let us now split the coordinates of ${\cal C}$ into
4-coordinates $X^\mu \equiv X^{1 \mu}~, ~~~~\mu=0,1,2,3$
associated with  position of a chosen particle, labeled by 1,
and the remaining coordinates $X^{\bar M}$:
\be
X^M \, = \,(X^\mu  ,\,X^{\bar M} )\,
\lbl{2.17}
\ee
The quadratic form occurring in the action (\ref{2.6}) can then
be split---according to the well known procedure of Kaluza-Klein
theories---into a 4-dimensional part plus the part due to the
extra dimensions of configuration space ${\cal C}$:
\be
\dot X^M \dot X^N \,G_{MN} \, = \,\dot X^\mu  \dot X^\nu  
g_{\mu \nu } \, + \,{\rm{extra}}\,{\rm{terms}}
\lbl{2.18}
\ee

More precisely, if for the metric of ${\cal C}$ we take the ansatz
\be
G_{MN} \, = \,\left( \begin{array}{l}
 g_{\mu \nu }  + A_\mu  ^{\bar M} A_\nu  ^{\bar N} \phi _{\bar M\bar N} 
 \,,\,\,\,\,\,A_\mu  ^{\bar N} \phi _{\bar M\bar N}  \\ 
 \,\,\,\,\,\,\,\,\,A_\nu  ^{\bar N} \phi _{\bar M\bar N} 
 \,,\,\,\,\,\,\,\,\,\,\,\,\,\,\,\,\,\,\,\,\,\,\,\,
 \phi _{\bar M\bar N} \,\,\,\,\,\,\,\,\,\,\, \\ 
 \end{array} \right)
\lbl{2.19}
\ee
then we obtain
\be
\dot X^M \dot X^N \,G_{MN} \, = \,\dot X^\mu  \dot X^\nu  g_{\mu \nu } \, + 
\,\dot X_{\bar M} \dot X_{\bar N} \,\phi ^{\bar M\bar N}
\lbl{2.20}
\ee
where
\be
\dot X_{\bar M} \, = \,G_{\bar MN} \dot X^N \, = \,A_{\bar M\mu } \dot X^\mu  
 + \phi _{\bar M\bar N} \dot X^{\bar N}
\lbl{2.21}
\ee

Inserting expression (\ref{2.20}) into the action (\ref{2.16}), we have
\be
I[X^\mu  ,X^{\bar M}] \, = \,M\int {d\tau \,\left[ {\dot X^\mu  \dot X^\nu  g_{\mu \nu }  + 
\,\phi ^{\bar M\bar N} (A_{\bar M\mu } \dot X^\mu   + \phi _{\bar M\bar J} \dot X^{\bar J} )
(A_{\bar N\nu } \dot X^\nu   + \phi _{\bar N\bar K} \dot X^{\bar K} )} \right]} ^{1/2}
\lbl{2.22}
\ee
where we have omitted subscript $m$.

Variation of the latter action with respect to $X^\mu$ gives
\be
\frac{1}{{( {\dot X^2 })^{1/2} }}\,\,\frac{d}{{d\tau }}
\left( {\frac{{\dot X^\mu  }}{{( {\dot X^2 })^{1/2} }}} \right)\, 
+ \,\,\frac{1}{{\dot X^2 }}\, \Gamma ^\mu  _{\rho \sigma } \dot X^\rho  \dot X^\sigma 
\, + \,{\rm{extra}}\,{\rm{terms}}\,{\rm{ = }}\,{\rm{0}}
\lbl{2.23}
\ee
where $\dot X^2 \, \equiv g_{\rho \sigma } \dot X^\rho  \dot X^\sigma $. 
This is just the 4-dimensional geodesic equation plus extra terms due to
the extra coordinates of  ${\cal C}$.

For explicit derivation it is convenient to use, instead of (\ref{2.16}), 
an equivalent action, namely the phase space action
\be
I\,\left[ {X^M ,P_M ,\Lambda } \right]\, = \,\int {d\tau } 
\left( {P_M \dot X^M \, - \,H} \right)
\lbl{2.24}
\ee
where
\be
H = \frac{\Lambda }{{2M}}\left( {P_M P_N \,G^{MN}  - M^2 } \right)
\lbl{2.25}
\ee
is the ``Hamiltonian" which---due to reparametrization invariance---is identically
zero. Variation of the action (\ref{2.24}) with respect to $P_M$ and
$\Lambda$, respectively, gives
\be
P_M  = \,\frac{{M\,\dot X_M \,}}{\Lambda }\,,\,\,\,\,\,\,\,\,\,\,\Lambda \, = 
\,\dot X_M \dot X_N \,G_{MN}
\lbl{2.25a}
\ee

Splitting variables $X^M$ according to (\ref{2.17}), and analogously for
$P_M$, we obtain
\be
I[X^\mu  ,X^{\bar M} ,p_\mu  ,P_{\bar M} ,\Lambda ] = 
\int {d\tau \,\left[ {p_\mu  \dot X^\mu   + 
P_{\bar M} \dot X^{\bar M} \, - \,H} \right]}
\lbl{2.26}
\ee
The Hamiltonian becomes
\be
H = \frac{\Lambda }{{2M}}\,\left[ {g^{\mu \nu } \left( {p_\mu   - 
A_\mu  ^{\bar J} P_{\bar J} } \right)\left( {p_\nu   - 
A_\nu  ^{\bar K} P_{\bar K} } \right) + \phi ^{\bar M\bar N} P_{\bar M} P_{\bar N} 
 - M^2 } \right]
\lbl{2.27}
\ee
where $p_\nu = P_\nu$ is 4-dimensional momentum.

Let us now assume that the ``internal" subspace of ${\cal C}$  admits
isometries given by the Killing vector fields $k_\alpha^{\bar J}$.
Index $\alpha$ runs over the independent Killing vectors, whereas
${\bar J}$, like ${\bar M},~{\bar N}$, runs over the ``internal" coordinates.
Then, as it is customary in Kaluza-Klein theories, we write
\be
A_\mu  ^{\bar J}  = k_\alpha  ^{\bar J} A_\mu  ^\alpha  
\lbl{2.28}
\ee
The metric $\phi^{{\bar M}{\bar N}}$ of the internal space
can be rewritten in terms of a metric
$\varphi^{\alpha \beta}$ in the space of isometries:
\be
\phi ^{\bar M\bar N}  = \varphi ^{\alpha \beta } k_\alpha  ^{\bar M} k_\beta  ^{\bar N}
+ \phi_{\rm extra}^{{\bar M}{\bar N}}
\lbl{2.29}
\ee
Here $\phi_{\rm extra}^{{\bar M}{\bar N}}$ are additional terms due to the directions
that are orthogonal to ismotries. For particular
internal spaces ${\bar {\cal C}}$, those additional terms may vanish.

Introducing projections of momentum onto Killing vectors
\be
p_\alpha  \, \equiv \,\,k_\alpha  ^{\bar J} P_{\bar J} 
\lbl{2.30}
\ee
and chosing a coordinate system in ${\cal C}$ in which
\be
k_\alpha  ^M  = \left( {k_\alpha  ^\mu  ,k_\alpha  ^{\bar M} } \right)\,,\,\,\,\,
k_\alpha  ^\mu   = 0,\,\,\,\,\,k_\alpha  ^{\bar M}  \ne 0
\lbl{2.30a}
\ee
Hamiltonian (\ref{2.27}) reads:
\be
H = \frac{\Lambda }{{2M}}\,\left[ {g^{\mu \nu } \left( {p_\mu   -
A_\mu^\alpha \, p_\alpha  } \right)\left( {p_\nu   
- A_\nu  ^\beta \, p_\beta  } \right) + \varphi ^{\alpha \beta } p_\alpha  p_\beta   - M^2 } \right]
\lbl{2.31}
\ee
For simplicity we will omit the extra terms $\phi_{\rm extra}^{{\bar M}{\bar N}}$.

Now we can use the Hamilton equations of motion:
\be
\dot p_\alpha   = \left\{ {p_\alpha  ,\,H} \right\}
\lbl{2.32}
\ee
\be
\dot p_\mu   = \left\{ {p_\mu  ,\,H} \right\}
\lbl{2.33}
\ee
Calculating the Poisson brackets
\be
\left\{ {p_\alpha  ,\,p_\beta  } \right\} = \frac{{\partial p_\alpha  }}{{\partial X^J }}\frac{{\partial p_\beta  }}{{\partial X_J }}\, - \,\frac{{\partial p_\beta  }}{{\partial X^J }}\frac{{\partial p_\alpha  }}{{\partial X_J }}\, = \,\left( {k_{\alpha ,J} ^M k_\beta  ^J  - k_{\beta ,J} ^M k_\alpha  ^J } \right)p_M  =  - \,C_{\alpha \beta } ^\gamma  p_\gamma  
\lbl{2.34}
\ee
introducing the kinetic momentum
\be
p_\mu   - A_\mu  ^{\bar J} P_{\bar J} \, \equiv \,\pi _\mu  \,,\,\,\,\,\,\,\,\,
g^{\mu \nu } \,\pi _\nu  \, = \,\frac{M}{\Lambda }\dot X^\mu  
\lbl{2.35}
\ee
and the gauge field strength
\be
F_{\mu \nu } ^\alpha   = \partial _\mu  A_\nu  ^\alpha   - 
\,\,\partial _\nu  A_\mu  ^\alpha   + C_{\alpha '\beta '} ^\alpha  A_\mu  ^{\alpha '} A_\nu  ^{\beta '} 
\lbl{2.35a}
\ee
we obtain
\be
\dot p_\alpha  \, = \,C_{\alpha \beta } ^\gamma  \,p_\gamma  \,A_\mu  ^\beta  \,\dot X^\mu  
\, - \frac{\Lambda }{{2M}}\varphi ^{\alpha '\beta '} _{,\bar J} p_{\alpha '} p_{\beta '} \,k_\alpha  ^{\bar J} 
\lbl{2.36}
\ee
\be
\dot \pi _\mu   - \frac{\Lambda }{{2M}}\,g_{\rho \sigma ,\mu } \pi ^\rho  \pi ^\sigma   + F_{\mu \nu } ^\alpha  p_\alpha  \,\dot X^\nu   
+ \frac{\Lambda }{{2M}}\,\left( {\varphi ^{\alpha \beta } _{,\mu }  - \,\varphi ^{\alpha \beta } _{,\bar J} \,k_{\alpha '} ^{\bar J} A_\mu  ^{\alpha '} } \right)p_\alpha  p_\beta  \, = \,0
\lbl{2.37}
\ee
This is the well known {\it Wong equation}\,\ci{Wong}, with additional terms due to the
presence of scalar fields $\varphi^{\alpha \beta}$.

\subsection{Relation between the higher dimensional and 4-dimensional mass}

If we rewrite the quadratic form (\ref{2.20}) as
\be
\frac{{\dot X^\mu  \dot X^\nu  g_{\mu \nu } }}{{\dot X^M \dot X^N \,G_{MN} }}\,\, = \,1\,\, - 
\,\,\frac{{\dot X_{\bar M} \dot X_{\bar N} \,\phi ^{\bar M\bar N} }}{{\dot X^M \dot X^N \,G_{MN} }}
\lbl{2.38}
\ee
and multiply by $M^2$, we find
\be
M^2 \frac{{\dot X^\mu  \dot X^\nu  g_{\mu \nu } }}{{\dot X^M \dot X^N \,G_{MN} }}\, = 
\,M^2 \, - \,\,\phi ^{\bar M\bar N} p_{\bar M} p_{\bar N} \, = \,g^{\mu \nu } p_\mu  p_\nu  \, = \,\,m^2 
\lbl{2.39}
\ee
where
\be
P_M  = \,\,\frac{{M\,\dot X_M }}{{\left( {\dot X^J \dot X^K \,G_{JK} } \right)^{1/2} }}
\lbl{2.40}
\ee
From eq.(\ref{2.39}) we obtain the ratio of the mass $m$ in $M_4$ to the mass $M$ in
${\cal C}$ expressed in terms of the corresponding velocity quadratic form:
\be
\frac{m}{M}\, = \,\,\left( {\frac{{\dot X^\mu  X^\nu  
g_{\mu \nu } }}{{\dot X^M \dot X^N \,G_{MN} }}} \right)^{1/2} 
\lbl{2.41}
\ee
For the 4-dimensional momentum we have
\be
P_\mu  \, = \,\,\frac{{M\,\dot X_\mu  }}{{\left( {\dot X^J \dot X^K \,G_{JK} } \right)^{1/2} }}
\,\, = \,\,\,\frac{{m\,\dot X_\mu  }}{{\left( {\dot X^\mu  \dot X^\nu  \,g_{\mu \nu } } 
\right)^{1/2} }}\,\, = \,p_\mu  
\lbl{2.42}
\ee
This is the same as eq.\,(\ref{2.12a}), but now derived for a more general metric
of ${\cal C}$.
Using eqs.(\ref{2.39}),(\ref{2.42}), together with (\ref{2.25a}),
the equation of motion (\ref{2.37}), after raising free indices, assume the
form
 $$\frac{1}{\lambda }\frac{d}{{d\tau }}\left( {\frac{{\dot X^\mu  }}{\lambda }} \right) + 
 \,^{(4)} \Gamma ^\mu  _{\rho \sigma } \frac{{\dot X^\rho  \dot X^\sigma  }}{{\lambda ^2 }}
 \,\, + \,\frac{{p_\alpha  }}{m}F_{\mu \nu } ^\alpha  \,\frac{{\dot X^\nu  }}{\lambda } $$
\be
 \,\,\,\,\,\,\,\,\,\,\,\,\,\,\,\,\,\,\,\,\,\,\,\,\,\,\,\, + \,\,\,\frac{1}{{2m^2 }}
 \,\left( {\varphi ^{\alpha \beta } _{,\mu }  - \,\varphi ^{\alpha \beta } _{,\bar J}
  \,k_{\alpha '} ^{\bar J} A_\mu  ^{\alpha '} } \right) p_\alpha  p_\beta 
   \, + \,\frac{1}{{\lambda m}}\,\frac{{\dd m}}{{\dd \tau }} = \,0 \\ 
\lbl{2.43}
\ee
where $\lambda = \left( {\dot X^\mu  \dot X^\nu  \,g_{\mu \nu } } \right)^{1/2}$.

From eq.\,(\ref{2.43}), in which $p_\alpha$ have the role of gauge charges, we see
that $m$ has the role of {\it inertial mass} in 4-dimensions. Four dimensional mass $m$
is given by higher dimensional mass $M$ and the contribution due to the extra
components of momentum $P_{\bar M}$:
\be
m^2 \, = \,g^{\mu \nu } p_\mu  p_\nu  \, = \,\,M^2 \, - 
\,\,\phi ^{\bar M\bar N} p_{\bar M} p_{\bar N} \, =
   M^2 - \varphi ^{\alpha \beta } \,p_\alpha  p_\beta  
\lbl{2.44}
\ee
These extra components $P_{\bar M}$ are in fact momenta of all other particles
within the configuration. In general $m$ is not constant, but in configuration spaces
with suitable isometries it may be constant.

A configuration under consideration can be the universe. Then, according to this theory,
the motion of a subsystem, approximated as a point particle, obeys the law of motion
(\ref{2.43}). Besides the usual 4-dimensional gravity, there are extra forces. 
They come from the generalized metric, i.e., the metric of configuration space. Since
the inertial mass of a given particle
depends on momenta of other particles and their states of motion (their momenta),
the Mach principle is automatically incorporated in this theory.
Such approach opens a Pandora's box of possibilities to revise our current
views on the universe. Persisting problems, such as the horizon problem,
dark matter, dark energy, the Pioneer effect, etc., can be examined afresh  within
this theoretical framework based on the concept of configuration space.

Locality, as we know it in the usual 4-dimensional relativity, no longer
holds in this new theory, at least not in general. But in particular,
when the metric of ${\cal C}$ assumes the block diagonal form (\ref{2.14}),
we recover the ordinary relativity (special and general), together
with locality. However, it is reasonable to expect that metric
(\ref{2.14}) may not be a solution of the Einstein equations in ${\cal C}$.
Then the ordinary relativity, i.e., the relativity in $M_4$, could
be recovered as an approximation only. Even before going
into the intricate work of solving the equations of general relativity
in ${\cal C}$, we already have a crucial prediction, namely that
locality in spacetime holds only approximately. When considering
the universe within this theory, we have to bear in mind that
the concept of spacetime has to be replaced by the concept of
configuration space ${\cal C}$. Locality in $M_4$ has thus to
be replaced by locality in ${\cal C}$. More technically this means
that, instead of differential equations in $M_4$ (e.g., the Einstein equations),
we have differential equations in ${\cal C}$: a given configuration
(a point in ${\cal C}$) can only influence a nearby configuration
(a nearby point in ${\cal C}$). Only in certain special cases
this translates into the usual notion of locality in $M_4$
(a subspace of ${\cal C}$). The so called `horizon problem'
does not arise in this theory.

\section{Strings, branes}

Theories of strings and higher dimensional objects---branes---are
very promising in explaining the origin and interrelationship
of the fundamental interactions, including gravity\,\ci{strings,branes} .

But there is a cloud. A question arises as to what is the
geometric principle behind string and brane theories, and
how to formulate them in a background independent way\,\ci{geometric}.

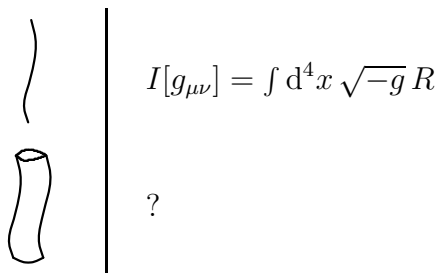
\begin{figure}[ht]
\setlength{\unitlength}{.4mm}
\begin{picture}(120,110)(-110,15)

\put(40,105){\line(0,-1){88}}
\put(53,78){$I[g_{\mu \nu}] = \int  \dd^4 x \, \sqrt{-g}\, R $}
\put(53,36){?}

\thicklines

\spline(14,67)(12,73)(15,82)(17,93)(15,101)

\spline(9,22)(7,28)(10,37)(12,48)(10,56)
\spline(19,22)(17,28)(19,37)(22,48)(20,56)
\closecurve(10,56, 12,54.5, 15,54.5, 17,55, 20,56, 17,57.5, 
        15,57.9, 12,57.5, 10,56)
\spline(9,22)(12,20.5)(14,20.3)(16,20.5)(19,22)

\end{picture}

\caption{ \small To point particle there corresponds the 
Einstein-Hilbert action
in spacetime. What is a corresponding space and action for a closed string?}

\end{figure}

Since such a fundemantal issue has been left unsettled
in the course of  the development of string theory, it
is not difficult to imagine that the latter theory
is not yet finished. Recent serious criticism of string
theory refers to an incomplete theory\ci{Smolin}. In the following
we will consider the possibility that string/brane theories
should take into account the concept of configuration
space. 

\subsection{Configuration space for infinite dimensional
objects -- branes}

A brane can be considered as a point in an infinite dimensional
space ${\cal M}$ with coordinates
\be
X^\mu  (\xi ^a )\, \equiv \,X^{\mu (\xi )} \, \equiv X^M 
\lbl{3.1}
\ee
where $X^\mu  (\xi ^a )$, $\mu=0,1,2,…,N-1;~a=0,1,3,…,n-1,~
n<N$, are brane's embedding fuctions\,\ci{PavsicBook,PavsicPortoroz}.
This includes classes
of tangentially deformed branes, which we can interpret
as being physically different objects, not just  as being
related by reparametrizations of the brane's world manifold
\ci{PavsicBook} (see Fig.\,3).

\setlength{\unitlength}{.8mm}

\begin{figure}[ht]
\hs{3mm} \begin{picture}(120,60)(25,0)
\put(10,0){\epsfig{file=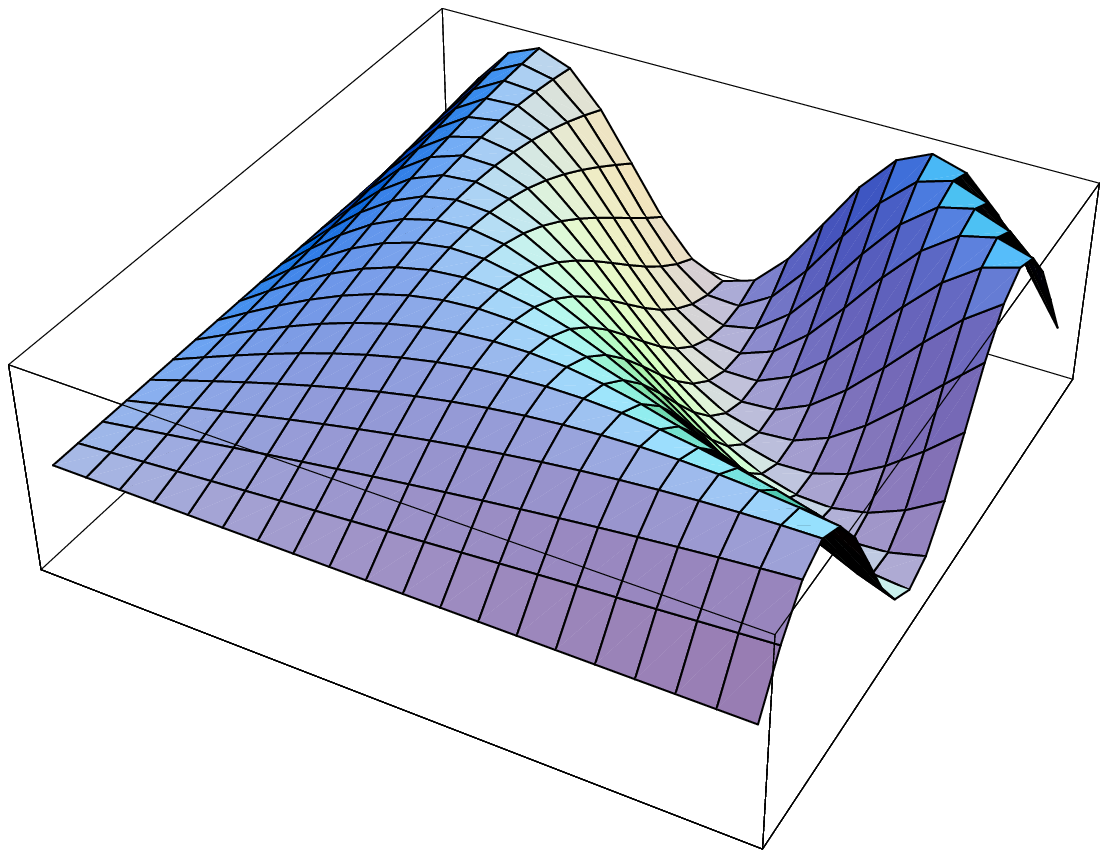,width=78mm}}
\put(98,0){\epsfig{file=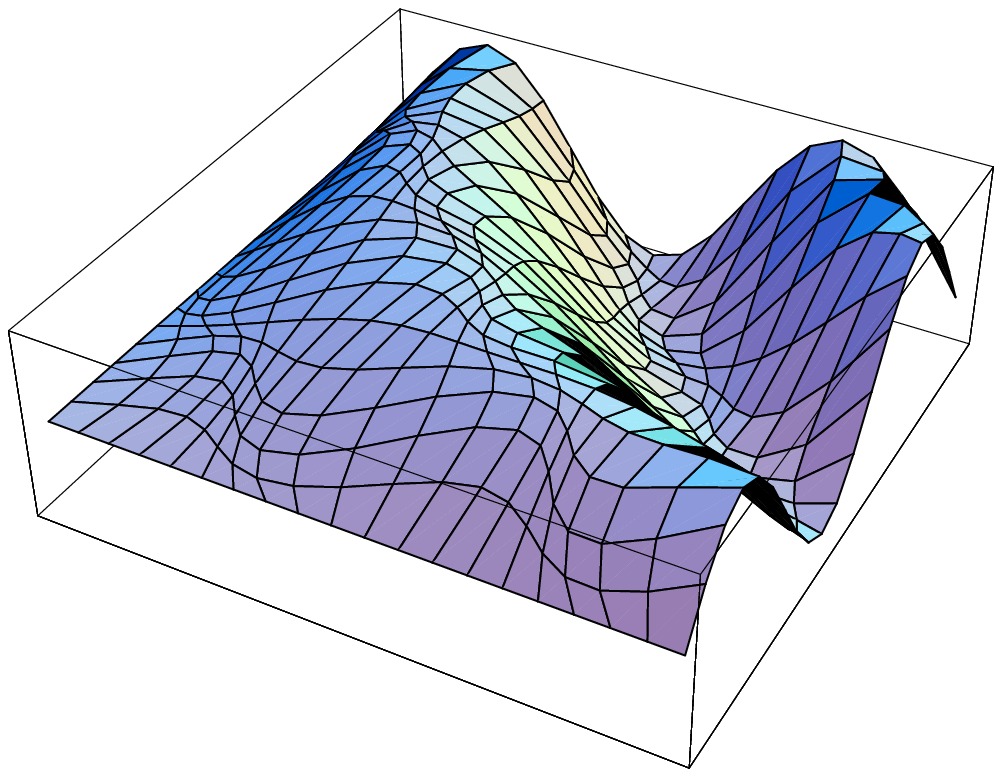,width=78mm}}
\end{picture}

\caption{\small Examples of tangentially deformed membranes. 
Mathematically the surface on the left is the same as the surface
on the right. Physically the two surfaces are different.}
\end{figure} 

All such objects are represented by different points of 
${\cal M}$-space. The latter space is the configuration
space associated with a brane. This is the space of all
(infinitely many) possible configurations of a brane.

Instead of one brane we can take a 1-parameter family of
branes $ X^\mu  (\tau,\xi ^a )\, \equiv \,X^{\mu (\xi )}(\tau)
 \, \equiv X^M (\tau) $, i.e., a curve (trajectory) in ${\cal M}$.
In principle every trajectory is kinematically possible.
A particular dynamical theory then selects which amongst
those kinematically possible branes and trajectories are
dynamically possible. We assume that dynamically possible
trajectories are {\it geodesics} in ${\cal M}$ determined by
the minimal length action \ci{PavsicBook,PavsicPortoroz}:
\be
I[X^M ] = \int {{\rm{d}}\tau } \,\,(\rho _{MN} \,\dot X^M \dot X^N )^{(1/2)} 
\lbl{3.2}
\ee
Here $\rho_{MN}$ is the metric of ${\cal M}$.

In particular, if metric is
\be
\rho _{MN} \, \equiv \,\rho _{\mu (\xi ')\nu (\xi '')} \, = 
\,\kappa \,\frac{{\sqrt {|f(\xi ')|} }}{{\sqrt {\dot X^2 \,(\xi ')} }}
\,\delta (\xi ' - \xi '')\,\eta _{\mu \nu }
\lbl{3.3}
\ee
where $f_{ab} \, \equiv \partial _a X^\mu  \partial _b X^\nu  \eta_{\mu \nu } $
is the induced metric on the brane,  $f \equiv \det \,f_{ab}$,
$\dot X^2  \equiv \dot X^\mu  \dot X^\nu  g_{\mu \nu } $, ($\eta_{\mu \nu}$
being the Minkowski metric of the embedding spacetime), then the equations
of motion derived from (\ref{3.2}) are precisely those of a Dirac-Nambu-Goto
brane \cite{PavsicBook,PavsicPortoroz}.

In this theory we assume that metric (\ref{3.3}) is just one particular
choice amongst many other possible metrics of ${\cal M}$. But
dynamically possible metrics are not arbitrary. We assume that
they must be solutions of the Einstein equations in ${\cal M}$
\cite{PavsicBook,PavsicPortoroz}. 

We take the brane space ${\cal M}$ as an arena for physics.
The arena itself is a part of the dynamical system,
it is not prescribed in advance.
The theory is thus background independent. It is based on the geometric
principle which has its roots in the brane space ${\cal M}$.

\begin{figure}[ht]
\setlength{\unitlength}{.4mm}
\begin{picture}(120,90)(-80,15)

\put(40,105){\line(0,-1){88}}
\put(53,78){$I[g_{\mu \nu}] = \int  \dd^4 x \, \sqrt{-g}\, R $}
\put(53,36){$I[\rho_{\mu (\phi) \nu (\phi')}] = \int {\cal D} X \, 
       \sqrt{|\rho|} \, {\cal R}$ }

\thicklines

\spline(14,67)(12,73)(15,82)(17,93)(15,101)

\spline(9,22)(7,28)(10,37)(12,48)(10,56)
\spline(19,22)(17,28)(19,37)(22,48)(20,56)
\closecurve(10,56, 12,54.5, 15,54.5, 17,55, 20,56, 17,57.5, 
        15,57.9, 12,57.5, 10,56)
\spline(9,22)(12,20.5)(14,20.3)(16,20.5)(19,22)

\end{picture}

\caption{ \small Brane theory is formulated in ${\cal M}$-space. The action
is given in terms of the ${\cal M}$-space curvature scalar ${\cal R}$.
We use abbreviation
$\phi \equiv \phi^A = (\tau,\xi^a)$.
}

\end{figure}

In summary, the infinite dimensional brane space ${\cal M}$ has in
principle any metric that is a solution to the Einstein's equations in
${\cal M}$. For the particular diagonal metric (\ref{3.3}) we obtain the ordinary
branes, including strings. But it remains to be checked whether such
particular metric is a solution of this generalized dynamical system
at all. If not, then this would mean that the ordinary string and brane
theory is not exactly embedded into the theory based on dynamical
${\cal M}$-space. The proposed theory goes beyond that 
of the usual strings and branes. It  resolves the problem
of background independence and the geometric principle behind the
string theory (Fig. 4). Geometric principle behind the string theory
is based on the concept of brane space ${\cal M}$, i.e., the configuration
space for branes. Occurrence of gauge and gravitational fields in
string theories is also elucidated. Such fields are due to string
configurations. They occur in the expansion of a string state
functional in terms of the Fock space basis. This can now be understood
as well within the classical string theory based on the action (\ref{3.2})
with ${\cal M}$-space metric $\rho_{MN}$, which is dynamical and
satisfies the Einstein equations in ${\cal M}$. Multidimensionality
of $\rho_{MN}$ allows for extra gauge interactions, besides gravity.
In the following we will point out how in the infinite dimensional
space ${\cal M}$ one can factor out a finite dimensional subspace.

\subsection{Finite dimensional description of extended objects}

The Earth has a huge (practically infinite) number of degrees of
freedom. And yet, when describing the motion of the Earth around
the Sun,  we neglect them all, except for the coordinates
of  {\it the centre of mass.}

Instead of infinitely many degrees of freedom associated with an
extended object, we may consider {\it a finite number of degrees
of freedom}. 

Strings and branes have infinitely many degrees of freedom. But at
first approximation we can consider just {\it the centre of mass}
(Fig.\,5a).

\setlength{\unitlength}{.8mm}

\begin{figure}[ht]
\hs{3mm} \begin{picture}(120,40)(10,-5)
\put(50,0){\epsfig{file=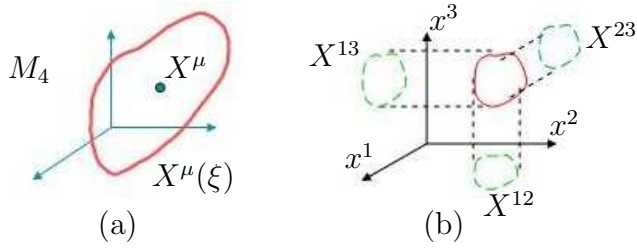,width=78mm}}

\put(50,20){$M_4$}
\put(76,20){$X^\mu$}
\put(74,2){$X^\mu(\xi)$}
\put(106,5){$x^1$}
\put(140,10.5){$x^2$}
\put(120,28){$x^3$}
\put(65,-6){(a)}
\put(119,-6){(b)}
\put(100,22){$X^{13}$}
\put(129,-3){$X^{12}$}
\put(146,25){$X^{23}$}

\end{picture}

\caption{\small With a closed string one can associate the centre
of mass coordinates (a), and the area coordinates (b)).  }
\end{figure}

Next approximation is in considering the holographic coordinates
$X^{\mu \nu}$ of the {\it oriented area} enclosed by the
string (Fig.\,5b).

We may go further and search for evenetual thickness of the object.
If the string has finite thickness, i.e., if actually it is not a string, but
a 2-brane, then there exist the corresponding volume degrees of
freedom $X^{\mu \nu \rho}$ (Fig.\,6).

\begin{figure}[ht]
\hs{3mm} \begin{picture}(120,40)(10,0)
\put(55,0){\epsfig{file=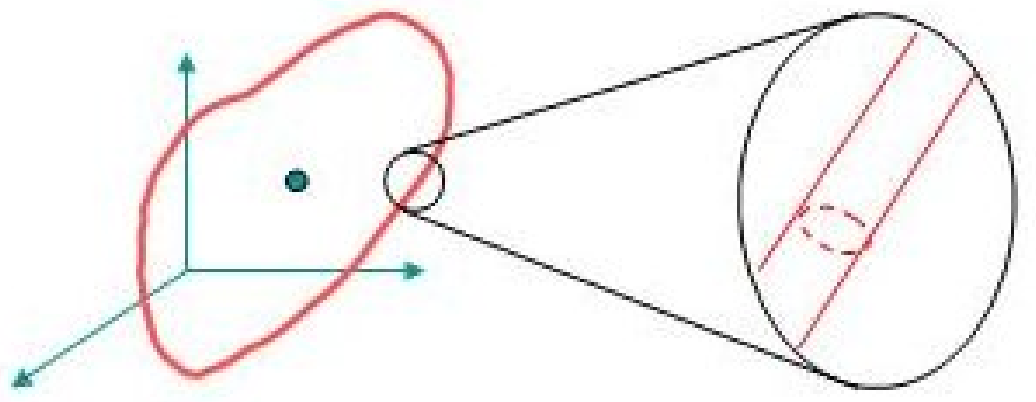,width=61mm}}

\put(50,20){$M_4$}
\put(74,19.5){$X^\mu$}
\put(74,1){$X^\mu(\xi)$}
\put(117,8){$X^{123}$}

\end{picture}

\caption{\small Looking with a sufficient resolution one can
detect eventual presence of volume degrees of freedom. }
\end{figure}

In general, for an extended object in $M_4$, we have 16 coordinates
\be
X^M \, \equiv \,X^{\mu _1 ...\mu _r } \,,\,\,\,\,\,\,\,r = 0,1,2,3,4
\lbl{3.4}
\ee
They are projections of $r$-dimensional volumes (areas) onto the
coordinate planes.

Oriented r-volumes can be elegantly described by Clifford
algebra\,\ci{Hestenes}. Instead of  the usual relativity, formulated
in spacetime in which the interval is
\be
{\rm{d}}s^2 \, = \,\,\eta _{\mu \nu \,} {\rm{d}}x^\mu  {\rm{d}}x^\nu  
\lbl{3.5}
\ee
one can consider the theory in which the interval is extended to
the space of $r$-volumes, called pandimensional continuum\,\ci{Pezzaglia} or
Clifford space\,\ci{Castro,PavsicBook,PavsicClifford}:
\be
{\rm{d}}S^2 \, = \,G_{MN} \,{\rm{d}}x^M {\rm{d}}x^N 
\lbl{3.6}
\ee
Coordinates of Clifford space can be used to model extended
objects \ci{Castro,PavsicArena}.
They are a generalization of the concept of center of mass.
Instead of describing an extended object in ``full detail, we
can describe it in terms of the center of mass, area and
volume coordinates. In particular, the extended object can be a
fundamental string/brane.

{\it Dynamics}  Taking also a time like parameter $\tau$, our object
can be described by 16 functions $X^M (\tau)$.
Let the action for an extended object described in terms of
the coordinates of Clifford space be
\be
I = \int \dd \tau \,(G _{MN} \dot X^M \dot X^N  )^{1/2}
\lbl{3.7}
\ee
If $G_{MN}=\eta_{MN}$ is Minkowski metric, then the equations
of motion are
\be
\ddot X^M \, \equiv \,\,\frac{{\,{\rm{d}}^{\rm{2}} X^M }}{{{\rm{d}}\tau ^2 }}
\,\, = \,\,0
\lbl{3.8}
\ee
They hold for tensionless branes. For the branes with tension one has to replace
$\eta_{MN}$ with a generic metric $G_{MN}$ with non vanishing curvature.
Eq.\,(\ref{3.8}) then generalizes to the corresponding geodesic equation
\be
     \frac{1}{\sqrt{{\dot X}^2} }\, 
     \left ( \frac{{\dot X}^M}{\sqrt{{\dot X}^2}}  \right )
     + \Gamma_{JK}^M \frac{{\dot X}^J {\dot X}^K}{{\dot X}^2} = 0
\lbl{3.9}
\ee     
Such higher dimensional configuration space, associated with branes, enables
unification of fundamental interactions \` a la
Kaluza-Klein\,\ci{PavsicKaluza}.
For alternative, although related approaches see \ci{CliffKaluzaAlternative}.

In quantizing such classical theory, based on the action (\ref{3.7})
which describes `point particle' in ${\cal C}$, one expects to be able
to proceed as usual, and arrive at the quantum field theory in ${\cal C}$.
So there would be no necessity to increase the dimensionality of
(4-dimensional) spacetime, from which we start in building the 16-dimensional
Clifford space ${\cal C}$.

Consider now the possibility of choosing to describe the object of Figs.\,5,6,
not with a set of coordinates
$X^M$ $=\{ X, X^{\mu_1},X^{\mu_1 \mu_2}$, $X^{\mu_1 \mu_2,\mu_3},
X^{\mu_1 \mu_2,\mu_3,\mu_4} \}$, but rather employ a more detailed
description. Let $\xi$ be a parameter along the ``centroid" loop,
i.e., a closed string of Fig.\,5a, and let $X^\mu (\xi)$ be its embedding
functions in $M_4$. But if we look more closely (Fig.\,6), we find that
at every value of parameter $\xi$ the string has a thickness: At every
value of $\xi$, instead of a point there may be a loop, described by
$X^{\mu_1 \mu_2}$. If we look even closer, we may find even more
structure, encoded in coordinates $X^{\mu_1 \mu_2,\mu_3}$ and
$X^{\mu_1 \mu_2,\mu_3,\mu_4}$. Altogether, taking into account
as well a time like parameter $\tau$, our object is described by 16
functions $X^M (\tau, \xi)$. These are functions describing the
embedding of a string's worldsheet into a 16-dimensional Clifford space
${\cal C}$. Instead of the usual string worldsheet $X^\mu (\tau, \xi)$,
$\mu=0,1,2,3$, we have a worldsheet in a higher dimensional space
which is not spacetime, but Clifford space (i.e., a configuration sapce)
${\cal C}$. It was
shown that such string living in ${\cal C}$---which happens to have
signature $(8,8)$---can be be consistently quantized\,\ci{PavsicSaasFee}
by employing the Jackiw definition of vacuum state
\,\ci{Jackiw,PavsicPseudoHarm}.

Hence, there is a possibility of having a consistent string theory
without employing extra dimensions of spacetime, provided that one
does not consider infinitely thin strings, but allows for their
thickness, encoded in the coordinates of Clifford space.

\section{Summary}

We have considered a theory in which spacetime is replaced by a larger space,
namely the configuration space ${\cal C}$ associated with a system under consideration.
The ordinary special and general relativity are recovered for particular classes of metric
of ${\cal C}$.  Since configuration space has extra dimensions,
its metric provides description of additional interactions, beside the
4-dimensional gravity, just as in Kaluza-Klein theories. They can occur
in {\it macro physics} and in {\it micro physics}. Eventual modification
of the dynamics at the levels from galaxies to the Universe is thus
based on the same underlying principle as the dynamics
of elementary particles (leaving quantum features aside).
In this theory there is no need for extra dimensions of {\it spacetime}.
The latter space is a subspace of the {\it configuration space} ${\cal C}$,
and all dimensions of ${\cal C}$ are physical.
Therefore, there is no need for a compactification of the extra dimensions of 
${\cal C}$.
 
We have presented some theoretical justifications for
the insight that the basic space of physics could be  associated with
configurations of physical systems.  Search for possible realistic solutions
to the theory and their comparison with observations is a project worth
of being pursued in the future.

\end{document}